\documentclass[10pt,twocolumn,showpacs,amsmath,amssymb,aps,prb,groupedaddress]{revtex4}

\usepackage{graphicx}
\usepackage{dcolumn}
\usepackage{bm}
\usepackage[cp1250]{inputenc}
\usepackage[usenames,dvipsnames]{color}
\usepackage{tabularx}

\newcommand{\veci}{{\mathbf i}}
\newcommand{\vecj}{{\mathbf j}}
\newcommand{\vecn}{{\mathbf f}}
\newcommand{\vecf}{{\mathbf f}}
\newcommand{\vecg}{{\mathbf g}}
\newcommand{\vecS}{{\mathbf S}}
\newcommand{\vecL}{{\mathbf L}}
\begin{document}

\title{Exchange couplings for Mn ions in CdTe: \\
validity of spin models for dilute magnetic II-VI semiconductors}

\author{Thorben Linneweber$^1$}
\author{J\"org B\"unemann$^2$}
\author{Ute L\"ow$^1$}
\author{Florian Gebhard$^3$}
\author{Frithjof Anders$^1$}
\affiliation{$^1$Lehrstuhl f\"ur Theoretische Physik II,
Technische Universit\"at Dortmund, D-44227 Dortmund, Germany}
\affiliation{$^2$Institut f\"ur Physik, BTU Cottbus-Senftenberg, D-03013 Cottbus, 
Germany}
\affiliation{$^3$Fachbereich Physik, Philipps-Universit\"at Marburg,
D-35032 Marburg, Germany}

\date{\today}

\begin{abstract}%
We employ density-functional theory (DFT) in the generalized gradient
approximation (GGA) and its extensions GGA+$U$ and GGA+Gutzwiller 
to calculate the magnetic exchange couplings 
between pairs of Mn ions substituting Cd in a CdTe crystal 
at very small doping.
DFT(GGA) overestimates the exchange couplings by a factor of three
because it underestimates the charge-transfer gap in Mn-doped II-VI semiconductors. 
Fixing the nearest-neighbor coupling $J_1$ to its experimental value
in GGA+$U$, in GGA+Gutzwiller,
or by a simple scaling of the DFT(GGA) results provides 
acceptable values for the exchange couplings
at 2nd, 3rd, and 4th neighbor distances in Cd(Mn)Te, Zn(Mn)Te, 
Zn(Mn)Se, and Zn(Mn)S.
In particular, we recover the experimentally observed relation
$J_4>J_2,J_3$. The filling of the Mn 3$d$-shell is not integer which puts the 
underlying Heisenberg description into question.
However, using a few-ion toy model
the picture of a slightly extended local moment emerges so
that an integer $3d$-shell filling 
is not a prerequisite for equidistant magnetization
plateaus, as seen in experiment.
\end{abstract}


\maketitle

\section{Introduction}
\label{sec:introduction}

The introduction of spin degrees of freedom 
in semiconductors leads to a variety of new phenomena, e.g.,
giant Zeeman splitting or giant Faraday rotation;~\cite{Furdyna,Jain,Kossut}
for recent studies of spin-diffusion and spin relaxation
of Mn-doped semiconductor heterostructures, 
see Refs.~[\onlinecite{PhysRevB.93.195307,PhysRevB.82.035211}] 
and references therein.
Over the last decades, the field of diluted magnetic semiconductors 
has attracted a lot of attention 
with the perspective of using the spin degree of freedom for electronic devices
(`spintronics').~\cite{Spintronics2002,Dietl1019,RevModPhys.86.187,Schaepers}
Apart from potential applications,
the description of magnetic ions in a semiconducting host material poses
an interesting but difficult problem in theoretical condensed matter physics.

Mn-doped II-VI semiconductors were among the first diluted magnetic 
semiconductors to be studied intensively.~\cite{Furdyna,Jain}
For small doping, the isovalent Mn ions replace the Cd ions.
Early on, it was pointed out that the Mn ions possess magnetic moments
whose couplings are mediated by the semiconductor host material.
The observation of equidistant magnetization steps
confirmed the assumption that the Mn ions carry spin $s=5/2$ and their mutual
interaction can be expressed in terms of a Heisenberg model with 
antiferromagnetic pair couplings~$J_n>0$
at $n$th-neighbor distance.~\cite{PhysRevB.33.1789,Foner1989,PhysRevLett.80.5425,Bindilatti2000,Bednarski2003172}

Not only the exchange
couplings between nearest neighbors but also those 
between Mn ions at 2nd, 3rd, and 4th neighbor distances were 
experimentally determined; all other couplings are negligibly small,
$J_{n\geq 5}\ll J_4$.
A modeling of the magnetization curves at very low temperatures
leads to the surprising result that
$J_4>J_2,J_3$,~\cite{PhysRevLett.80.5425,Bindilatti2000}
i.e., the exchange couplings do not decay monotonously as a function
of the geometrical distance.

The unexpected non-monotonous decay of $J_n$ as a function of the Mn-Mn 
separation, and also the overall size of the exchange couplings,
are unexplained.
Only the nearest-neighbor 
exchange couplings $J_1$ for Mn-doped II-VI semiconductors
were calculated using the superexchange 
approach,~\cite{PhysRevB.33.3407,PhysRevB.37.4137,refId0,PhysRevB.90.075205} 
or density-functional theory (DFT) 
in the local-density approximation, DFT (LSDA), and
in DFT(LSDA)+$U$.~\cite{PhysRevB.79.205204,PhysRevB.83.239903}

In this work, we calculate the exchange couplings $J_{n\leq 4}$ using three
itinerant-electron approaches, (i),
the generalized-gradient approximation (GGA)
to DFT with the functional of Perdew, Burke, and Ernzerhof,~\cite{PBE},
(ii) GGA+$U$ as implemented in the FLEUR package,~\cite{FLEUR}
and, (iii), GGA+Gutzwiller for a suitable two-ion Hubbard model.
We confirm that $J_4>J_2,J_3$ and find a reasonable agreement with measured values
for Cd(Mn)Te, Zn(Mn)Te, Zn(Mn)Se, and Zn(Mn)S.
Furthermore, our analysis shows that the filling 
of the Mn $3d$-shell is not integer which
challenges the notion of Mn ions carrying a spin
$s=5/2$.
We study a few-ion toy model to show that 
the  non-integer filling remains consistent with
equidistant magnetization plateaus.
The picture of a spatially distributed spin $s=5/2$
emerges which includes the neighboring Wannier orbitals 
that hybridize with the Mn $3d$-states.
We therefore conclude that 
the concept of interacting Heisenberg spins remains applicable
for Mn ions diluted in II-VI semiconductors.

Our work is organized as follows. 
In Sect.~\ref{sec:supercell} we specify the setup for
our DFT(GGA) and GGA+$U$ calculations.
Moreover, we derive the two-ion Hubbard model for our GGA+Gutzwiller
approach, and define the exchange couplings in terms of 
ground-state energy differences of the itinerant electron description.
In Sect.~\ref{sec:results} we provide the exchange couplings
for up to fourth neighbors in Cd(Mn)Te, Zn(Mn)Te, Zn(Mn)Se, and Zn(Mn)S,
and compare them to experiment.
As an example, we discuss the magnetization as a function of magnetic field
for Cd$_{1-x}$Mn$_x$Te for very low doping, $x=0.005$.
In Sect.~\ref{sec:toymodels} we discuss the magnetization curve
for a few-ion toy model and show that equidistant 
magnetization plateaus can be observed even though 
the filling of the Mn $3d$-shell is not integer.
Short conclusions, Sect.~\ref{sec:conclusions}, close our presentation.

\section{Ion pairs in a semiconductor host}
\label{sec:supercell}

We are interested in the properties of manganese atoms
diluted in a II-VI host semiconductor at low temperatures
and in sizable magnetic fields. 
To be definite, we shall focus on CdTe.
For very small Mn concentrations~$x$
in Cd$_{1-x}$Mn$_x$Te, we may safely assume that the Mn$^{2+}$ ions 
substitute the isovalent Cd$^{2+}$ ions.
We tested that it is a reasonable approximation
to neglect lattice distortions in the theoretical analysis
because structural relaxations turned out to be small within the
DFT(GGA) calculations.

\begin{figure}[ht]
\begin{center}
\includegraphics[width=1.0\columnwidth]{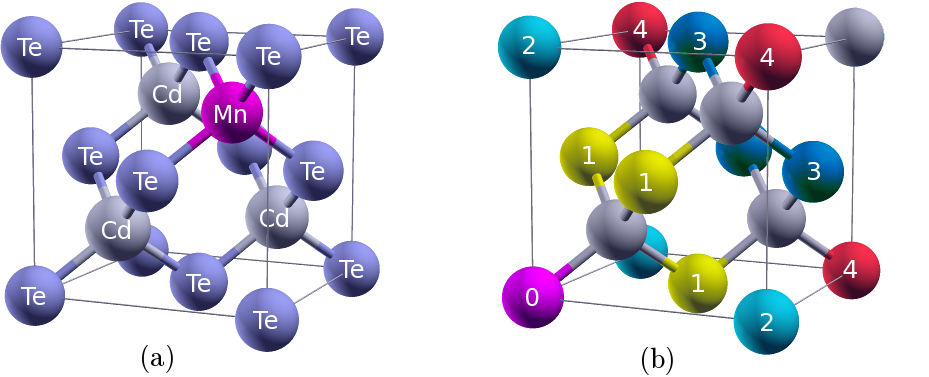}
\caption{(a) Zn-blende structure for Cd$_{0.75}$Mn$_{0.25}$Te where one out of four
Cd atoms is replaced by a Mn atom.\label{fig:fccunitcell}
(b) Positions of first, second, third, and fourth neighbors
on the Cd fcc sub-lattice from the Cd atom in the lower left corner. }
\end{center}
\end{figure}

CdTe crystallizes in the zinc-blende ($\beta$-ZnS) structure
where the fcc lattice of the Te ions is shifted against the fcc lattice
of the Cd ions by $a/4$ along the diagonal of the cubic cell of length
$a=6.482\, \hbox{\AA}$.~\cite{AshcroftMermin}
Fig.~\ref{fig:fccunitcell}(a) shows a fcc unit cell with one Mn atom 
replacing one out of four Cd atoms ($x=0.25$). 

The spin of an isolated Mn ion aligns with
any finite magnetic field. The non-trivial magnetization
curves seen in experiment are due to the exchange interaction
between different Mn ions.
Test calculations confirmed that 
the interaction of three or more Mn ions is given by the sum
of pair interactions so that we can concentrate on the
interaction between pairs of Mn ions as a function of their distance.
We found in numerically expensive
calculations with $L=512$ atoms in the unit cell that
the interaction between two Mn ions beyond 4th-neighbor
distance is negligibly small. 
In Fig.~\ref{fig:fccunitcell}(b) we show the first, second, third, and fourth 
neighbors on the fcc sublattice in CdTe.

\subsection{GGA and GGA+$\bm{U}$ calculations}
\label{sec:bandstructure}

Ideally, we should study a single pair of Mn ions
with Cd ions on all other sites of the cation fcc lattice.
However, practical band-structure calculations require translational symmetry. 
Therefore, we start from large but finite cells with $L$~atoms 
that contain two Mn ions, and link them together 
so that periodic boundary conditions apply in all three
spatial directions. Modern bandstructure program packages 
permit the investigation of large cells (`supercells').
In this work, we use supercells with $L\leq 128$ atoms
which are sufficiently large to study Mn pairs 
that are maximally fourth-nearest neighbors.

For our investigations we use the FLEUR package,~\cite{FLEUR}
a high-precision implementation of
the full potential linearized augmented planewave (FLAPW) approach
to density-functional theory (DFT) in the generalized gradient approximation (GGA).
The program package FLEUR also offers the option to include
the effect of the correlations between the electrons in the partly filled
$3d$-shell of the Mn ions on a mean-field level (GGA+$U$). 
In Sect.~\ref{sec:results}, we compile results for the
Mn-Mn interaction from both bandstructure approaches.

We run the FLEUR code using the following settings.
We use the Generalized Gradient Approximation (GGA)
functional of Perdew, Burke, and Ernzerhof~\cite{PBE}
for the exchange-correlation energy. Since we
are investigating a band insulator with a sizable gap,
it is sufficient to use only 10 inequivalent ${\bf k}$-points
in the irreducible part of the Brillouin zone; 
depending on the impurity positions, 
this corresponds to 20 or 40 ${\bf k}$-points in the full Brillouin zone.
The basis functions inside the muffin tins are expanded in spherical harmonic
functions with a cut-off of $L_{\rm max}=10$. The muffin tin radii are 
$R_{\rm Cd}=R_{\rm Mn}=2.64\, {\rm a.u.}$ and
$R_\mathrm{Te}=2.58\,\mathrm{a.u.}$
($1\, {\rm atomic\, unit} = a_{\rm B}=0.529\, \hbox{\AA}$).
We use $R_{\rm Te}K_{\rm max}=8.26$, where $K_{\rm max}$ 
is the plane wave cut-off. For the GGA+$U$
calculations we use the standard double counting correction.~\cite{Anisimov}

\subsection{GGA+Gutzwiller approach}
\label{sec:LDA+G}

The electrons in the Mn ions' $3d$ shell are strongly correlated.
Therefore, more sophisticated many-particle techniques should be
employed. For example, it would be desirable
to use the fully self-consistent Gutzwiller-DFT.~\cite{1367-2630-16-9-093034,Schmalian,Chinesen}
At present, however, the required large unit cells 
prevent us from doing such a calculation and we restrict ourselves
to a less costly method that is based on the evaluation
of a Gutzwiller wave function for a tight-binding model
with Hubbard-type interactions on the two Mn sites.

\subsubsection{Derivation of the two-ion Hubbard model}
\label{sec:deriveHubbardmodel}

The code {\sc Wannier90}
permits a downfolding of the bandstructure to a tight-binding 
Hamiltonian in position space.~\cite{Mostofi2008685}  
We project onto a basis of $s$~orbitals and $p$~orbitals for each
of the semiconductor atoms and $s$, $p$ and $d$~orbitals for the Mn impurity.
However, the downfolding procedure is
limited to $L=16$ atoms in the
unit cell so that we cannot derive the tight-binding model
for a pair of Mn ions directly.

To overcome this limitation,
we assume that the combined influence of two Mn ions 
on the electron transfer between two lattice sites can be
approximated by the linear superposition of the influence of two
individual Mn ions.
Under this linearity assumption, 
we are left with the investigation of a single Mn ion 
in a CdTe supercell of $L=16$ atoms.
For our GGA calculations we use 120 ${\bf k}$-points 
in the irreducible part of the Brillouin zone 
(1/24 of the full Brillouin zone) and $R_\mathrm{Te}K_{\rm max}=9.80$.

First, we calculate the bandstructure for pure CdTe.
The downfolding provides the tight-binding Hamiltonian for CdTe,
\begin{equation}
  \hat{H}^{\rm CdTe} = \sum_{\veci,\vecj,b_1,b_2,\sigma}
 t_{\veci,b_1,\sigma}^{\vecj,b_2,\sigma}
  \hat{c}_{\veci,b_1\sigma}^{\dagger}
\hat{c}_{\vecj,b_2\sigma}^{\vphantom{\dagger}} \; ,
\end{equation}
where $  \hat{c}_{\veci,b\sigma}^{\dagger}$ 
($\hat{c}_{\veci,b\sigma}^{\vphantom{\dagger}}$) creates (annihilates)
an electron in the orbital~$b$ with spin $\sigma=\uparrow,\downarrow$. 
Due to the symmetry of our crystal, there are no local hybridization terms,
and we may write
\begin{eqnarray}
  \hat{H}^{\rm CdTe} &=& \hat{T}^{\rm CdTe}+ \hat{V}^{\rm CdTe} \; , \nonumber \\
\hat{T}^{\rm CdTe}&=&\sum_{\veci\neq \vecj,b_1,b_2,\sigma}
 t_{\veci,b_1,\sigma}^{\vecj,b_2,\sigma}
  \hat{c}_{\veci,b_1\sigma}^{\dagger}
\hat{c}_{\vecj,b_2\sigma}^{\vphantom{\dagger}} \; ,\nonumber \\
\hat{V}^{\rm CdTe} &=& 
\sum_{\veci,b,\sigma} t_{\veci,b,\sigma}^{\veci,b,\sigma}  \hat{n}_{\veci,b\sigma}
\; ,
\end{eqnarray}
where $\hat{n}_{\veci,b\sigma}=  \hat{c}_{\veci,b\sigma}^{\dagger}
\hat{c}_{\veci,b\sigma}^{\vphantom{\dagger}} $
counts the number of electrons with spin $\sigma$ in orbital~$b$
on site~$\veci$.

\begin{figure}[ht]
\begin{center}
\includegraphics[width=\columnwidth]{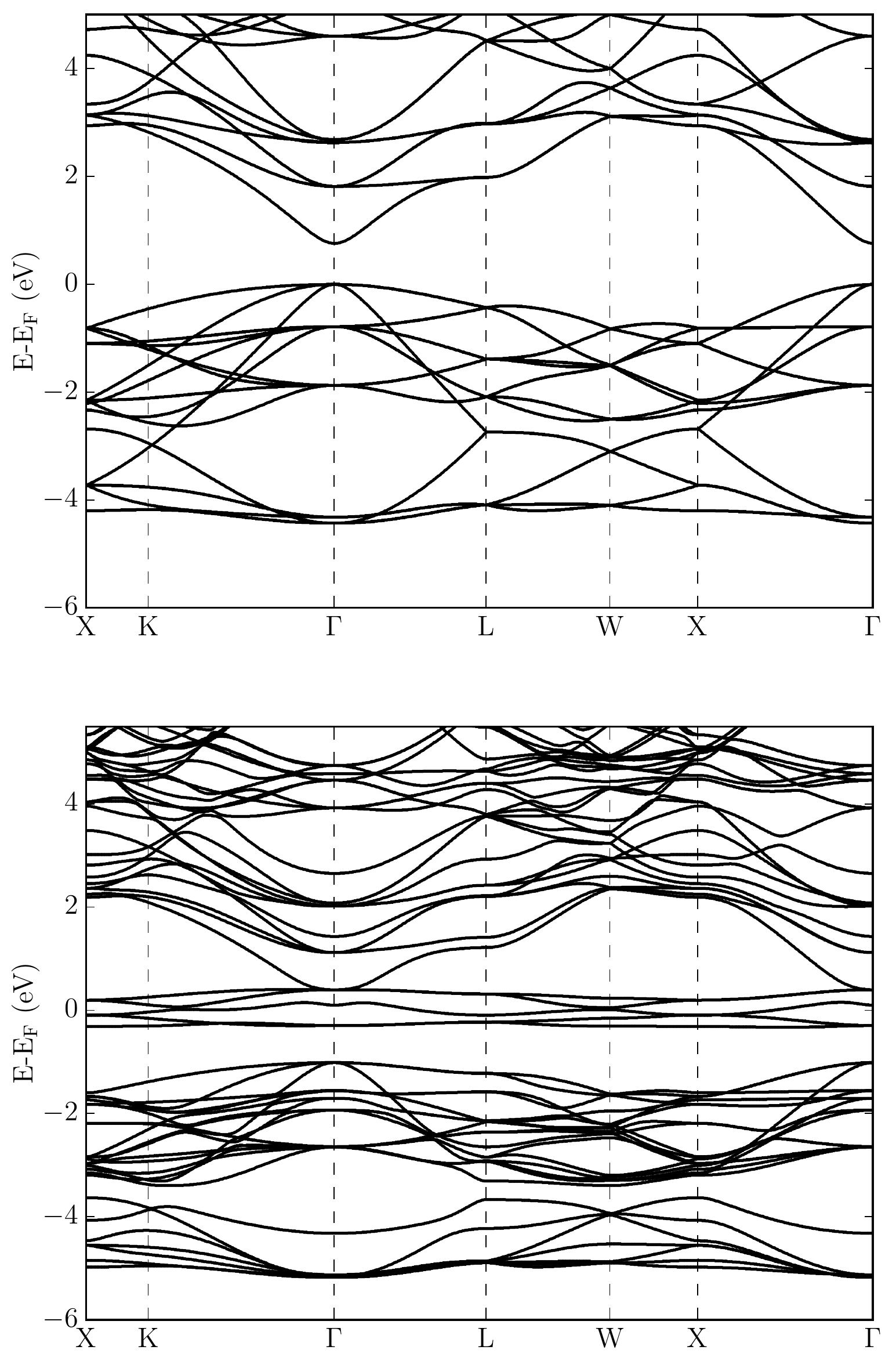}
\caption{Band structures of CdTe (upper part of the figure) 
and of Cd$_{0.875}$Mn$_{0.125}$Te
(lower part of the figure), calculated from a supercell with $L=16$ atoms without
(with)
a single Mn atom in the supercell using the FLEUR program 
package.\label{fig:CdTebandstructure}}
\end{center}
\end{figure}

Next, we repeat the calculation with a single Mn ion at position $\vecf$
in the supercell which leads to a new set of electron transfer matrix elements
$(t_{\veci,b_1,\sigma_1}^{\vecj,b_2,\sigma_2})^{\vecf}$,
\begin{equation}
  \hat{H}^{\rm CdTe, \vecf} = \sum_{\veci\neq\vecj,b_1,b_2,\sigma}
  (t_{\veci,b_1,\sigma}^{\vecj,b_2,\sigma})^{\vecf}
  \hat{c}_{\veci,b_1\sigma}^{\dagger}
\hat{c}_{\vecj,b_2\sigma}^{\vphantom{\dagger}}+\hat{V}^{\rm CdTe} \; .
\end{equation}
Due to periodic boundary conditions and the translational invariance of the
crystal, the bands for CdTe with a single Mn ion do not depend on~$\vecf$.
The corresponding bands for pure CdTe and with a single Mn ion 
in the $L=16$ supercell are shown in Fig.~\ref{fig:CdTebandstructure}.

The upper part of the figure shows that the direct gap at the $\Gamma$-point 
is $E_{\rm sp}=0.76\, {\rm eV}$, in agreement with 
previous calculations.~\cite{Wu201518}
However, DFT(LDA) and DFT(GGA) underestimate the
gap for the insulator CdTe. The (exciton) gap, a lower bound on the
single-particle gap, is found at $E_{\rm xc}=1.5\, {\rm eV}$
for CdTe.~\cite{ThomasgapCdTe}
DFT(GGA) also underestimates the charge-transfer gap 
in Mn-doped II-VI semiconductors between Te and Mn levels
so that the resulting exchange couplings are too large, 
see Sect.~\ref{sec:couplingstrengths}.
The origin of the exchange coupling can be inferred
from the lower part of Fig.~\ref{fig:CdTebandstructure}. 
The Mn $3d$-bands are grouped around the
Fermi energy so that they push down the CdTe bands that were below the gap,
and hybridize at the $\Gamma$-point with a dominant Te-band 
above the CdTe gap. The band structure shows that
the Mn-Te hybridization is responsible for the interaction between 
Mn ions.

To set up our Hamiltonian in the presence of two Mn impurities, 
we define the corrections to the electron transfer amplitudes
\begin{equation}
\bigl(\Delta_{\veci,b_1,\sigma}^{\vecj,b_2,\sigma}\bigr)^{\vecf}=
\bigl(t_{\veci,b_1,\sigma}^{\vecj,b_2,\sigma}\bigr)^{\vecf}-
t_{\veci,b_1,\sigma}^{\vecj,b_2,\sigma} \; .
\end{equation}
According to our linear superposition scheme, 
we model the presence of a second Mn impurity in our tight-binding Hamiltonian
by adding independently the corrections 
for the first Mn ion at site $\vecf_1$ and the second impurity 
at site~$\vecn_2$.
This defines our tight-binding Hamiltonian for the two-site problem,
{\arraycolsep=2pt\begin{eqnarray}
  \hat{H}^{\rm tb} &=& \sum_{\veci\neq\vecj,b_1,b_2,\sigma}
\bigl( t_{\veci,b_1,\sigma}^{\vecj,b_2,\sigma}\bigr)^{\vecf_1,\vecn_2}
\hat{c}_{\veci,b_1\sigma}^{\dagger}
\hat{c}_{\vecj,b_2\sigma}^{\vphantom{\dagger}}
+\hat{V}^{\rm CdTe}
\; ,
\nonumber \\
\bigl( t_{\veci,b_1,\sigma}^{\vecj,b_2,\sigma}\bigr)^{\vecf_1,\vecn_2}
&=&
\bigl(\Delta_{\veci,b_1,\sigma}^{\vecj,b_2,\sigma}\bigr)^{\vecf_1}+
\bigl(\Delta_{\veci,b_1,\sigma}^{\vecj,b_2,\sigma}\bigr)^{\vecn_2}+
t_{\veci,b_1,\sigma}^{\vecj,b_2,\sigma} \; .
\end{eqnarray}}%
Our approximation neglects the joint influence of the impurities
on the electron transfer matrix elements in their surrounding,
in the spirit of standard alloy theory.~\cite{Elliot}  
The supercells for the two-ion Hubbard model
can be much larger than those used for its construction ($L=16$).
For our further Gutzwiller calculations we work with cells containing $L=512$ atoms.

As a last step, we add the Hubbard interaction on the 
two Mn sites $\vecf_1$ and $\vecf_2$
and obtain the two-ion Hubbard model,
\begin{eqnarray}
\hat{H}&=&\hat{H}^{\rm tb} +\hat{H}^{\rm loc}_{\vecf_1}
+\hat{H}^{\rm loc}_{\vecf_2}
+ \hat{H}^{\rm dc} \nonumber \\
\hat{H}^{\rm loc}_{\vecg}&=&
\sum_{c_1,\ldots,c_4=1}^5\sum_{\sigma.\sigma'=\uparrow,\downarrow}
U^{(c_1\sigma),(c_2\sigma')}_{(c_3\sigma'),(c_4\sigma)}\nonumber \\
&& \hphantom{+ \sum_{c_1,\ldots,c_4=1}^5}
\hat{c}_{\vecg,c_1\sigma}^{\dagger}
\hat{c}_{\vecg,c_2\sigma'}^{\dagger}
\hat{c}_{\vecg,c_3\sigma'}^{\vphantom{\dagger}}
\hat{c}_{\vecg,c_4\sigma}^{\vphantom{\dagger}} \; ,\nonumber \\
\hat{H}^{\rm dc} &=& -E_{\rm dc}(  \hat{n}_{\vecf_1}+ \hat{n}_{\vecf_2}) \; .
\label{eq:Hamiltonian}
\end{eqnarray}
Here, $U^{.,.}_{.,.}$
describes the Coulomb interaction
between the electrons in the $3d$-shell in the ten spin-orbit level
$(c_l\sigma_l)$ in either of the two
Mn ions. Using some simplifying assumptions,
all interaction
coefficients can be expressed in terms of an intra-orbital
Hubbard-$U$ and an inter-orbital Hund's-rule~$J$,~\cite{1367-2630-16-9-093034}
see Sect.~\ref{sec:results}.

Lastly, $\hat{H}^{\rm dc}$ accounts for the double counting
of interaction terms between the $3d$-electrons
on a mean-field level,
where $\hat{n}_{\vecg}=\sum_{c,\sigma}\hat{n}_{\vecg,c\sigma}$
counts the number of correlated electrons on the Mn site~$\vecg$.
We use a particularly simple form for the double counting
term. In essence, the choice of $E_{\rm dc}$ permits to fix the average number of
electrons in the correlated Mn $3d$-orbitals and we shall present our 
results as a function of $n_d\equiv \sum_{c,\sigma} \langle
n_{\vecf_1,c,\sigma}\rangle
=\sum_{c,\sigma} \langle n_{\vecf_2,c,\sigma}\rangle $.
Typically, we need $E_{\rm dc}\approx 20\, {\rm eV}$ to adjust $n_d\approx 5$.

\subsubsection{Gutzwiller variational state}

We approximate the true ground-state 
of our model Hamiltonian~(\ref{eq:Hamiltonian})
by a Gutzwiller variational state,
\begin{equation}
| \Psi_{\rm G}\rangle= \hat{P}_{\vecf_1}\hat{P}_{\vecf_2} | \Phi_0\rangle \; ,
\label{eq:productAnsatz}
\end{equation}
where $| \Phi_0\rangle$ is the ground state of an (effective) single-particle 
Hamiltonian $H_{0}^{\rm qp}$, and $\hat{P}_{\vecg}$ is
the single-site Gutzwiller correlator,
\begin{equation}
\hat{P}_{\vecg} = \sum_{\Gamma} \lambda_{\Gamma} \hat{m}_{\vecg,\Gamma}
\label{eq:diagonal}
\end{equation}
with  $\vecg=\vecf_1,\vecf_2$.
Here, $\hat{m}_{\vecg,\Gamma}=
|\Gamma\rangle_{\vecg} {}_{\vecg}\langle \Gamma|$ projects onto 
the atomic eigenstate $|\Gamma\rangle_{\vecg}$ of $\hat{H}^{\rm loc}_{\vecg}=
\sum_{\Gamma}E_{\Gamma}\hat{m}_{\vecg,\Gamma}$, and $\lambda_{\Gamma}$
are real variational parameters for each of the $2^{10}=1024$ states in the
Mn $3d$-shell.

\subsubsection{Gutzwiller approximation and energy minimization}

To obtain the optimal values of the
variational parameters $\lambda_{\Gamma}$ and
the optimal single-particle product state $|\Phi_0\rangle$,
we must minimize the energy functional
\begin {equation}
E^{\rm var}(\{ \lambda_{\Gamma},|\Phi_0\rangle\})
= \frac{\langle \Psi_{\rm G} | \hat{H} | \Psi_{\rm G}\rangle}{
\langle \Psi_{\rm G} | \Psi_{\rm G}\rangle}  \; .
\label{eq:Evar}
\end{equation}
We evaluate the expectation value in eq.~(\ref{eq:Evar})
using the Gutzwiller approximation.~\cite{1367-2630-16-9-093034}
This corresponds to a neglect of 
correlations between the two Mn impurity sites.

Due to the presence of a second Mn impurity, the point group 
on each Mn site is not exactly cubic. Hence, the local
density matrix for the correlated orbitals 
\begin{equation}
C_{\vecg,c\sigma,c'\sigma} = 
\langle \Phi_0 |
\hat{c}_{\vecg,c'\sigma}^{\dagger}
\hat{c}_{\vecg,c\sigma}^{\vphantom{\dagger}} 
| \Phi_0 \rangle 
\end{equation}
is not diagonal. However, the non-diagonal elements are very small,
of the order of $10^{-3}$, and are therefore neglected in our
calculations, i.e., we set
\begin{equation}
C_{\vecg,c\sigma,c'\sigma} = \delta_{c,c'}n_{c,\sigma}  \; .
\label{eq:localdens}
\end{equation}
For the same reason, we use the approximation that 
the matrix for the electron transfer renormalization is diagonal,
$q_{c\sigma}^{c'\sigma'}=\delta_{c,c'}\delta_{\sigma,\sigma'}q_{c,\sigma}$.
Then, the energy functional can be cast into the form~\cite{PhysRevB.57.6896}
\begin {eqnarray}
E^{\rm GA}(\{ \lambda_{\Gamma},|\Phi_0\rangle\}) 
&=& \langle \Phi_0 | \hat{T}  | \Phi_0\rangle  
+ \sum_{\veci,b,\sigma} t_{\veci,b,\sigma}^{\veci,b,\sigma}  
\langle \Phi_0 | \hat{n}_{\veci,b\sigma}  | \Phi_0\rangle  \nonumber \\
&& + \sum_{\Gamma} E_{\Gamma} \lambda_{\Gamma}^2
\langle \Phi_0 |( \hat{m}_{\vecf_1,\Gamma}+ \hat{m}_{\vecn_2,\Gamma})
| \Phi_0\rangle 
\nonumber \\
&& -E_{\rm dc}
\langle \Phi_0 |  \hat{n}_{\vecf_1}+ \hat{n}_{\vecn_2} | \Phi_0\rangle  \; ,
\label{eq:energyGA}
\end{eqnarray}
where
\begin{equation}
\hat{T} = \sum_{\veci\neq \vecj,b_1,b_2,\sigma}
q_{b_1,\sigma}q_{b_2,\sigma}
\bigl( t_{\veci,b_1,\sigma}^{\vecj,b_2,\sigma}\bigr)^{\vecf_1,\vecn_2}
  \hat{c}_{\veci,b_1\sigma}^{\dagger}
\hat{c}_{\vecj,b_2\sigma}^{\vphantom{\dagger}} \; .
\end{equation}
The $q$-factors depend on the variational parameters $\lambda_{\Gamma}$
and the local densities $n_{c,\sigma}$; explicit expressions can be found in
Ref.~[\onlinecite{1367-2630-16-9-093034}].
We include the Lagrange parameter $E_{\rm SP}$
for the normalization of $|\Phi_0\rangle$
and $\eta_{c,\sigma}$ to fulfill eq.~(\ref{eq:localdens}).
Then, the minimization of the energy functional~(\ref{eq:energyGA})
with respect to $\langle \Phi_0 |$ 
leads to the effective single-particle problem~\cite{PSSB:PSSB201147585}
\begin{eqnarray}
\hat{H}_0^{\rm qp} |\Phi_0 \rangle &=& E_{\rm SP} | \Phi_0\rangle \nonumber \; ,\\
\hat{H}_0^{\rm qp} &=& \hat{T} + \sum_{\veci,b,\sigma}
t_{\veci,b,\sigma}^{\veci,b,\sigma}  
\hat{n}_{\veci,b\sigma}  \nonumber \\
&& -\sum_{c,\sigma} (E_{\rm dc} +\eta_{c,\sigma})
\Bigl(\hat{n}_{\vecf_1,c\sigma}+\hat{n}_{\vecn_2,c\sigma}\Bigr) \; .
\end{eqnarray}
The Lagrange parameters $\eta_{c,\sigma}$ are variational parameters
that control the local spin density in the $3d$-levels on the Mn ions,
while the double-counting energy $E_{\rm dc}$ determines
the average number of Mn $3d$ electrons.

\subsection{Exchange couplings}
\label{sec:heismodel}

The notion of an `exchange coupling'  between the two Mn atoms
hinges on the concept of a Heisenberg exchange between the two Mn 
impurity spins at $\vecf_1$ and $\vecn_2$,
\begin{equation}
\hat{H}_{\rm Heis}^{\vecf_1,\vecf_2}= 
2J_{\vecf_1-\vecf_2}\vecS_{\vecf_1} \cdot \vecS_{\vecf_2}
\; .
\label{eq:Heisenbergmodel}
\end{equation}
Here, we tacitly assume that the average 
filling of the $3d$-shell in the Mn atoms
is close to integer filling, i.e., $n_d\approx 5$,
and the Hund's-rule coupling fixes the ground-state spin to
$s=5/2$ on each ion. The exchange coupling is positive, 
$J_{\vecf_1-\vecn_2}>0$,
for an antiferromagnetic coupling.

Under the assumption that a Heisenberg model provides an adequate
description of the ground state (and low-energy excitations) of our 
two Mn impurities, we can estimate
their exchange coupling using the bandstructure and GGA+Gutzwiller
approach.
We orient the Mn spins into the $z$-direction, either parallel 
(`ferromagnetic alignment')
or antiparallel (`Ne\'el-antiferromagnetic alignment').
The algorithm converges to the corresponding (local) minima
and provides $(\Delta E)_{\vecf_1-\vecf_2}=E_{\parallel}-E_{\perp}$ for the 
energy differences.
This energy difference can also be calculated from the Heisenberg 
model~(\ref{eq:Heisenbergmodel}),
\begin{eqnarray}
(\Delta E)_{\vecf_1-\vecf_2}&=&2J_{\vecf_1-\vecf_2}
\langle {\rm FM} |\vecS_{\vecf_1} \cdot \vecS_{\vecf_2}| {\rm FM}\rangle
\nonumber \\
&& - 2J_{\vecf_1-\vecf_2}
\langle {\rm AFM} |\vecS_{\vecf_1} \cdot \vecS_{\vecf_2}| {\rm AFM}\rangle
\nonumber \\
&=&4J_{\vecf_1-\vecf_2} (5/2)^2= 25 J_{\vecf_1-\vecf_2} \; ,
\end{eqnarray}
with the spin states $|{\rm FM}\rangle= |5/2,5/2\rangle_{\vecf_1}
|5/2,5/2\rangle_{\vecf_2}$ and
$|{\rm AFM}\rangle= |5/2,5/2\rangle_{\vecf_1}|5/2,-5/2\rangle_{\vecf_2}$. Here
we used that 
only the $z$-components contribute to the expectation values.
In this way, the values $J_{\vecf_1-\vecf_2}=(\Delta E_{\vecf_1-\vecf_2})/25$
are accessible from approaches
that employ itinerant electrons. 

\section{Results}
\label{sec:results}

First, we show that the experimentally observed exchange couplings 
for Mn ion pairs up to 4th-neighbor distance can be 
reproduced from scaled DFT(GGA), GGA+$U$, and GGA+Gutzwiller.
Second, we analyze the local occupancies 
as obtained from GGA+Gutzwiller.

\subsection{Exchange couplings}

The values for the exchange couplings $J_n$ in Cd(Mn)Te are known from experiment
for up to 4th neighbors on the cation fcc lattice.
The values for the couplings have been determined from the steps
in the magnetization as a function of the externally applied field
for very low temperatures, $T\lesssim 0.1\, {\rm K}$.
Their sequence, e.g., the fact that $J_4>J_2, J_3$, has been extracted
from a fit of the data to cluster spin models. 
Bindilatti et al.~\cite{Bindilatti2000} find
$J_1=6.1\pm 0.3\, {\rm K}$,
$J_2=0.06\pm 0.01\, {\rm K}$,
$J_3=0.18\pm 0.01\, {\rm K}$, and
$J_4=0.39\pm 0.02\,{\rm K}$.
In this section we derive and compare the exchange couplings
from DFT(GGA), GGA+$U$ and GGA+Gutzwiller calculations, and
compare the resulting magnetization curves with experiment.

\begin{table}[t]
\begin{tabular}[t]{|c|r|rrrr|}
\hline
$J$ & \vphantom{\Large $A^a_a$}Exp. & GGA & s$\cdot$GGA & GGA+$U$ & GGA+G \\
\hline
\vphantom{\Large $A^a_a$}$J_1$ & 6.1\hphantom{0} & 17.1\hphantom{0} & 
6.1\hphantom{0}& 6.1\hphantom{0} & 6.1\hphantom{0}\\
$J_2$ & \hphantom{0}0.06 & 0.30  & 0.11  & 0.10 & 0.10\\
$J_3$ & \hphantom{0}0.18 & 0.96 & 0.34  & 0.30 & 0.27\\
$J_4$ & \hphantom{0}0.39 & 1.44 & 0.51 & 0.49 & 0.61\\
\hline
\end{tabular}
\caption{Heisenberg exchange couplings $J_n$ in~K
between Mn ions at $n$th neighbor distance on the Cd fcc lattice in {\bf CdTe}
from experiment,~\cite{Bindilatti2000} and from DFT(GGA),
DFT(GGA) scaled by a factor $s=0.357$, 
GGA+$U$ for $\overline{U}=U-J=2.65\, {\rm eV}$,
and GGA+Gutzwiller for $A=4.4\, {\rm eV}$, $B=0.1\, {\rm eV}$, $C=0.4\, {\rm eV}$
and $n_d=5.19$. \label{tab:Jvalues}}
\end{table}

\subsubsection{Coupling strengths}
\label{sec:couplingstrengths}

The DFT(GGA) calculation does not contain any specific 
parameters to adjust the exchange couplings. 
For large supercells, $L=128$, 
the influence of Mn pairs between neighboring supercells
in negligibly small. 

As seen from table~\ref{tab:Jvalues},
the value for the nearest-neighbor coupling from DFT(GGA)
is too large by more than a factor of two,
$J_1^{\rm DFT}=17.1\, {\rm K}\approx J_1/0.36$.
DFT(GGA) overestimates 
the size of the exchange coupling because
it finds a too small charge-transfer gap
$\Delta_{\rm CT}$
between occupied Te levels and unoccupied Mn-levels in Cd(Mn)Te.
In super-exchange models,~\cite{PhysRevB.90.075205}
the exchange integral $J_1$ is inversely proportional
to $\Delta_{\rm CT}$ so that the exchange integral $J_1$
becomes too large in DFT(LDA) and DFT(GGA), by almost a factor of three.
GGA+$U$ is frequently used to tackle gap problems 
in correlated insulators. When we apply a Hubbard-$U$ on the Mn sites,
we find a larger charge-transfer gap which leads to smaller 
exchange couplings, see below. As mentioned in Sect.~\ref{sec:bandstructure},
the gap in pure CdTe is too small in DFT(GGA) calculations.
This can also be corrected using GGA+$U$.~\cite{Wu201518}
However, the exchange couplings between Mn ions are mediated by electron transfer
processes between Mn and Te so that the precise value of the
CdTe band gap is irrelevant for our considerations.

\begin{figure}[t]
\begin{center}
\includegraphics[width=1.0\columnwidth]{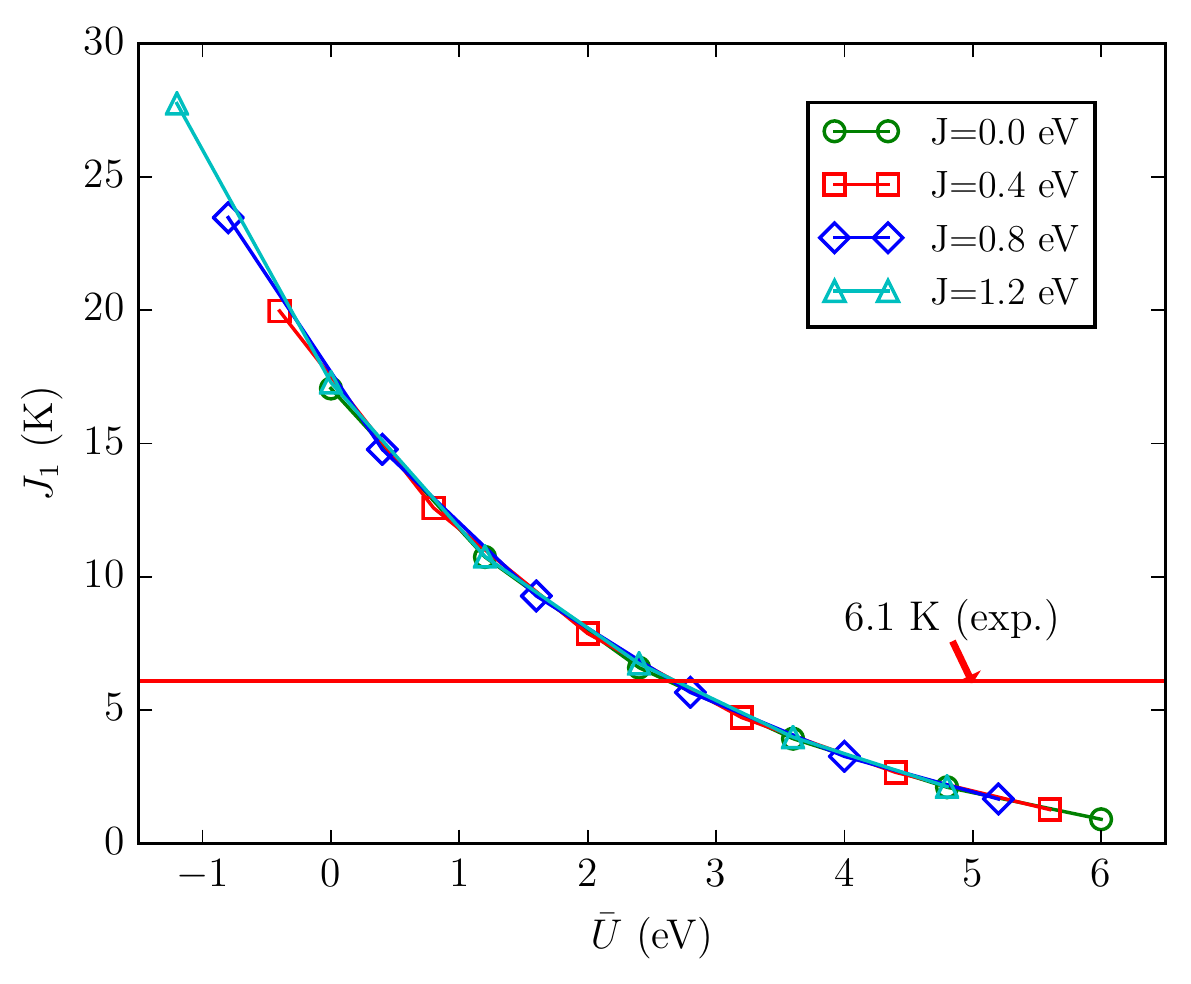}
\caption{Heisenberg exchange coupling $J_1$ between two Mn ions in CdTe 
at nearest-neighbor distance from GGA+$U$ as a function of
$\overline{U}=U-J$ calculated for a supercell with $L=128$ atoms
using the FLEUR program package. 
The red horizontal line shows the experimental value $J_1^{\rm exp}=6.1\, {\rm K}$.
\label{fig:JfromLDA+U}}
\end{center}
\end{figure}

In Fig.~\ref{fig:JfromLDA+U}
we show the dependence of $J_1^{{\rm GGA}+U}$ as a function of $U$
for various values of~$J$. The exchange coupling only depends
on the combination $\overline{U}=U-J$.~\cite{PhysRevB.57.1505}
For $\overline{U}=2.65\, {\rm eV}$ we obtain $J_1^{{\rm GGA}+U}=6.1\, {\rm K}$.
The values for other exchange interactions for farther distances are collected
in table~\ref{tab:Jvalues}. The values for $J_{2,3,4}^{{\rm GGA}+U}$
are very similar, and even slightly closer to experiment,
than those from the scaled DFT(GGA). This demonstrates that
an adjustment of the charge-transfer gap cures in effect 
the overestimation of the exchange interactions in DFT(GGA).

Lastly, we discuss the results for $J_n$ as obtained 
from our GGA+Gutzwiller calculations. 
We set $C=0.4\, {\rm eV}$, in agreement
with crystal-field theory for data from infrared spectroscopy 
for isolated Mn$^{2+}$ ions in CdTe.~\cite{Jain}
Moreover, we use $C=4B$, i.e., $B=0.1\, {\rm eV}$,
as is a reasonable assumption for transition metals.~\cite{Sugano1970}
A similar set of values was used in a recent study of exchange integrals
in Mn-doped II-VI semiconductors.~\cite{PhysRevB.90.075205}
The Hubbard-parameter~$U$ in transition metals is of the order of several 
eV.~\cite{PhysRevB.74.125106} In this work we set
$A=4.4\, {\rm eV}$.
Note that we have $U=A+4B+3C$ and $J=(5/2)B+C$
for our intra-orbital Hubbard interaction and Hund's-rule coupling,
or, for the Slater-Condon parameters,
we have $F^{(0)}=A+(7/5)C$, $F^{(2)}=49B+7C$, 
and $F^{(4)}=(63/5)C$.~\cite{ironpaper}
Therefore, our Hund's-rule exchange on the Mn sites is $J=0.65\, {\rm eV}$
and we employ $F^{(0)}=4.96\, {\rm eV}$ or $U=6\, {\rm eV}$.

\begin{figure}[ht]
\begin{center}
\includegraphics[width=1.0\columnwidth]{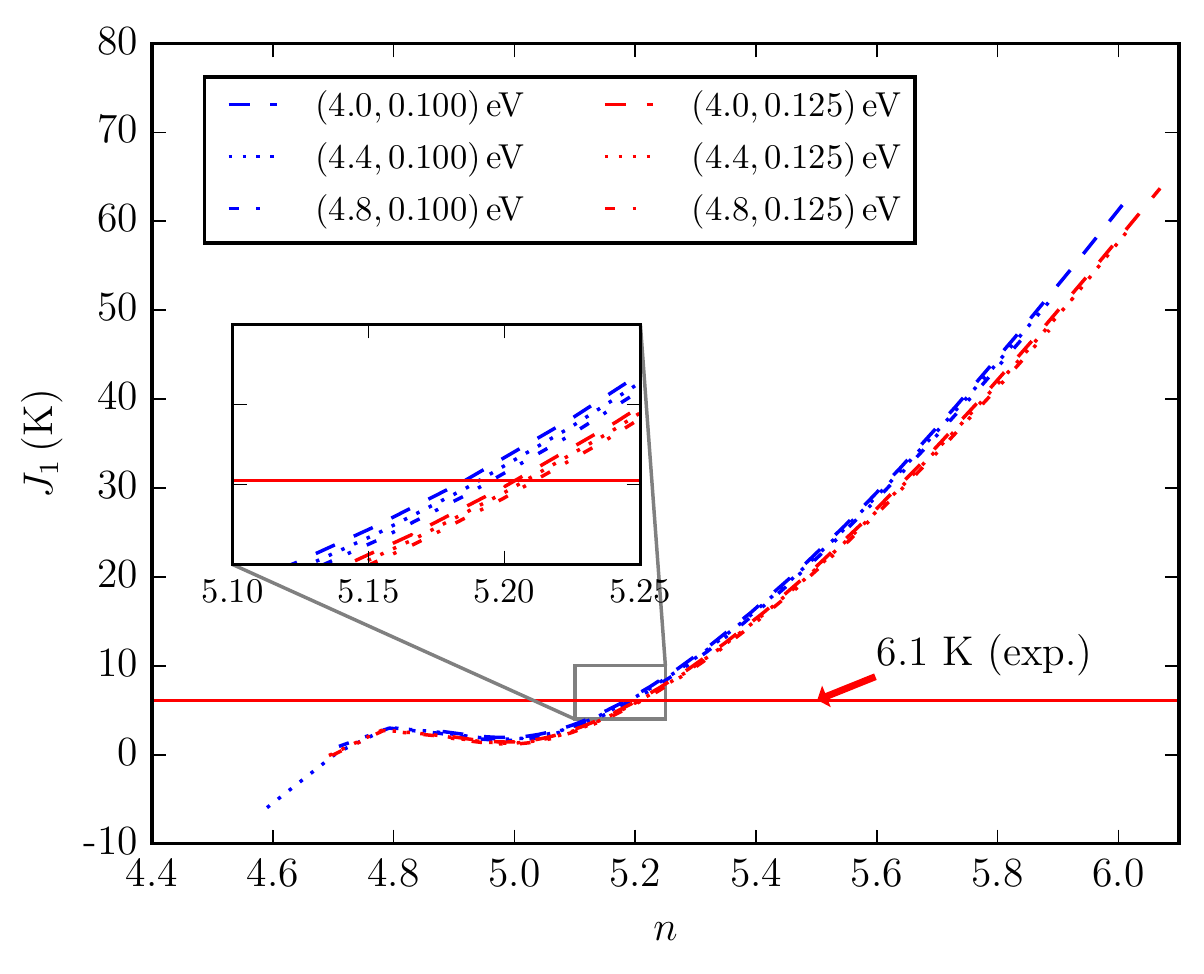}
\caption{Heisenberg exchange coupling $J_1$ between two Mn ions in CdTe 
at nearest-neighbor distance from GGA+Gutz\-wil\-ler as a function of 
the number of electrons in the Mn $3d$-shell for various values of
the Racah parameters $A$, $B$, and $C=4B$.
The red horizontal line shows the experimental value, $J_1^{\rm exp}=6.1\, {\rm K}$.
\label{fig:JfromGGA+Gutz}}
\end{center}
\end{figure}

In Fig.~\ref{fig:JfromGGA+Gutz},
we show $J_1$ as a function of the electron number~$n_d$ 
in the Mn $3d$-shell.
As seen from the figure,
the curves for $4.0\, {\rm eV} \leq A \leq 4.8\, {\rm eV}$ and
$0.3\, {\rm eV} \leq C \leq 0.5\, {\rm eV}$ essentially collapse onto each other 
in the region of interest, $J_1 =6.1\, {\rm K}$. Therefore, the specific choice of
the Racah parameters is not crucial.
As also seen from Fig.~\ref{fig:JfromGGA+Gutz}, 
the filling is not integer. Instead, we find that $n_d= 5.19$
reproduces the nearest-neighbor Heisenberg exchange coupling best
for $A=4.4\, {\rm eV}$, $B=0.1\, {\rm eV}$, $C=0.4\, {\rm eV}$.
The resulting values for the exchange couplings for Mn ions
in CdTe are compiled in table~\ref{tab:Jvalues}.

Our Gutzwiller calculations 
here are very close to a Hartree-Fock calculation.
Correlation effects are small for the two fully polarized Mn atoms with 
their (anti-) parallel spins. 
We discuss this point further in Sect.~\ref{sec:localocc}.
This agreement is specific for Mn in II-VI semiconductors
because we encounter a fully polarized, half-filled $3d$ shell 
in a wide-gap insulator. In other systems, correlation
effects are more pronounced, as seen in some preliminary calculations
for Cr-doped CdTe or Mn-doped GaAs.

For future reference, we compile the exchange couplings 
for Zn(Mn)Se, Zn(Mn)Te, and Zn(Mn)S in tables~\ref{tab:three}, \ref{tab:two},
and~\ref{tab:four}.
Note that the exchange couplings for $n>4$ are at least
an order of magnitude smaller than $J_2,J_3$, of the order
of $J_{n\geq 5}=0.01\, {\rm K}$, or less. 
This justifies our restriction to $J_{n\leq 4}$.

As seen from the tables, the GGA+Gutzwiller method
overestimates by some 20\%-30\% the nearest-neighbor exchange couplings~$J_1$
for Zn-VI semiconductors (VI=Te, Se, S) when we use 
$A=4.4\, {\rm eV}$, $B=0.1\, {\rm eV}$, $C=0.4\, {\rm eV}$
and $n_d=5.19$ for the Mn ions. 
With this parameter set, the method can be used to
provide a reasonable estimate for the nearest-neighbor couplings
for Mn ion pairs in II-VI semiconductors. 
GGA+Gutzwiller provides a much better estimate for the couplings $J_{n\geq 2}$ 
than DFT(GGA) but they are still systematically 
too large by a factor two to three.

\subsubsection{Magnetization for small doping 
and low temperatures}

As an application, we calculate the magnetization 
$M(B)$ as a function of the applied external field~$B$ for
Cd$_{1-x}$Mn$_x$Te at small but finite doping $x=0.005$.
A Mn ion is placed in the center of a large but finite fcc 
lattice with $50^3$ sites.
Then, Cd atoms in the surrounding of the `seed site' are replaced by Mn atoms
with probability~$x$.
As a first possibility, the central Mn ion remains isolated, i.e., with only
Cd atoms on its 1st, 2nd, 3rd, and 4th neighbor shell 
(`maximal surrounding'). In the absence of spin-orbit coupling,
the spin of such an isolated Mn ion aligns with
any finite magnetic field so that
its magnetic response is given by the Brillouin function.
Note that neglecting the spin-orbit coupling is justified because
of the full magnetic polarization of the Mn ions.~\cite{preprintPRB}

A second possibility are two-spin clusters with exactly one Mn ion
in the maximal surrounding of the seed site.
Two such clusters are equivalent when they can be mapped onto each other
by applying some space-group transformations of the fcc lattice.
Since equivalent clusters lead to the same magnetic response we only
need to store one representative~$C$ and determine its
multiplicity $A_C$.
Moreover, we need to calculate the probability~$p_C$
that a lattice point is part of cluster~$C$.~\cite{PhysRevB.54.6457,PhysRevLett.80.5425}
For example, for a nearest-neighbor cluster we have $A_{C}=12$
and $p_{C}=x^2(1-x)^{72}$ 
(because in this case 72 sites must be unoccupied).
This construction principle
is readily generalized for clusters with three or more spins.

\begin{table}[t]
\begin{tabular}[t]{|c|r|rrr|}
\hline
$J$ & \vphantom{\Large $A^a_a$}Exp. & GGA & s$\cdot$GGA & GGA+G \\
\hline
\vphantom{\Large $A^a_a$}$J_1$ & \hphantom{0}9.0\hphantom{0} 
& 41.2\hphantom{0} & 9.0\hphantom{0} & 11.45 \\
$J_2$ & 0.20 & 0.96  & 0.21 & 0.49 \\
$J_3$ & 0.16 & 2.61 & 0.57 & 0.54 \\
$J_4$ & 0.51 & 3.97 & 0.87 & 1.13  \\
\hline
\end{tabular}
\caption{Heisenberg exchange couplings $J_n$ in~K
between Mn ions at $n$th neighbor distance on the Zn fcc lattice in {\bf ZnTe}
from experiment,~\cite{PhysRevLett.80.5425} from (scaled) DFT(GGA),
and from GGA+Gutzwiller 
for $A=4.4\, {\rm eV}$, $B=0.1\, {\rm eV}$, $C=0.4\, {\rm eV}$
and $n_d=5.19$. \label{tab:three}}
\begin{tabular}[t]{|c|r|rrr|}
\hline
$J$ & \vphantom{\Large $A^a_a$}Exp. & GGA & s$\cdot$GGA & GGA+G \\
\hline
\vphantom{\Large $A^a_a$}$J_1$ & 12.2\hphantom{0} & 48.1\hphantom{0}  
&12.2\hphantom{0} & 14.97\\
$J_2$ & 0.16 & 0.81 & 0.21 &  0.28\\
$J_3$ & 0.07 & 1.61 & 0.41 &  0.42\\
$J_4$ & 0.43 & 3.26 & 0.82 &  1.16\\
\hline
\end{tabular}
\caption{Heisenberg exchange couplings $J_n$ in~K
between Mn ions at $n$th neighbor distance on the Zn fcc lattice in {\bf ZnSe}
from experiment,~\cite{PhysRevLett.80.5425} from (scaled) DFT(GGA),
and from GGA+Gutzwiller 
for $A=4.4\, {\rm eV}$, $B=0.1\, {\rm eV}$, $C=0.4\, {\rm eV}$
and $n_d=5.19$. \label{tab:two}}
\begin{tabular}[t]{|c|r|rrr|}
\hline
$J$ & \vphantom{\Large $A^a_a$}Exp. & GGA & s$\cdot$GGA & GGA+G \\
\hline
\vphantom{\Large $A^a_a$}$J_1$ & 16.9\hphantom{0} & 60.3\hphantom{0} 
& 16.9\hphantom{0} & 19.73 \\
$J_2$ & 0.27 & 0.99  &  0.28 & 0.47 \\
$J_3$ & 0.04 & 1.14 &  0.32 & 0.40 \\
$J_4$ & 0.41 &  2.85 &  0.80  &  0.97 \\
\hline
\end{tabular}
\caption{Heisenberg exchange couplings $J_n$ in~K
between Mn ions at $n$th neighbor distance on the Zn fcc lattice in {\bf ZnS}
from experiment,~\cite{PhysRevLett.80.5425} from (scaled) DFT(GGA),
and from GGA+Gutzwiller 
for $A=4.4\, {\rm eV}$, $B=0.1\, {\rm eV}$, $C=0.4\, {\rm eV}$
and $n_d=5.19$. \label{tab:four}}
\end{table}

In this work we include clusters with one to four Mn atoms
and thus find in total 1130 inequivalent clusters~$C$. At 
doping $x=0.005$, clusters with up to three Mn atoms
cover 98.5\%~of all possible configurations,
clusters with up to four Mn atoms
cover 99.6\% of all possible configurations.
Therefore, clusters with five and more Mn atoms
are irrelevant at $x=0.005$.

\begin{figure}[t]
\begin{center}
\includegraphics[width=0.9\columnwidth]{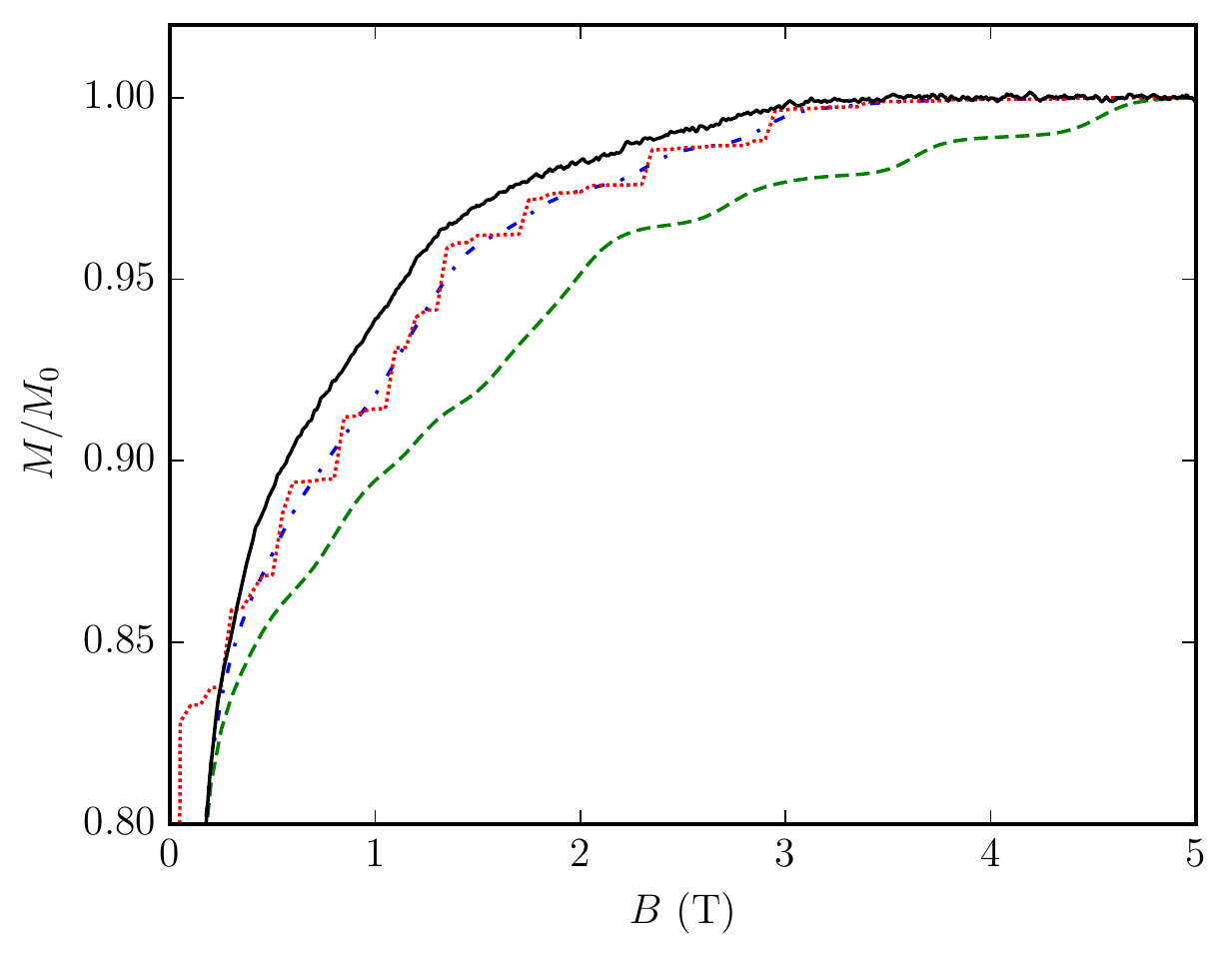}
\caption{Magnetization~$M(B)$ as a function of the external field~$B$
normalized to its value at $B=5\, {\rm T}$ for 
Cd$_{1-x}$Mn$_x$Te at Mn-doping  $x=0.005$. 
Black line (solid): experimental curve;~\cite{Bindilatti2000}
red line (dotted): Heisenberg model with experimental coupling parameters 
at $T=0$;
blue line (dash-dotted): Heisenberg model with experimental coupling parameters 
at $T_{\rm spin}=0.1\, {\rm K}$;
green line (dashed): Heisenberg model with GGA+Gutzwiller
parameters at $T_{\rm spin}=0.1\, {\rm K}$.
Clusters with up to four Mn ions are included.\label{fig:bindilattimag}}
\end{center}
\end{figure}

For each cluster~$C$, the interaction between the Mn spins 
is described by a Heisenberg model, 
\begin{equation}
\hat{H}_{\rm Heis}^C(B)
= \sum_{\substack{\vecf_1,\vecf_2\in C\\(\vecf_1\neq{}\vecf_2)}}
J_{\vecf_1-\vecn_2}\hat{\vecS}_{\vecf_1} \cdot \hat{\vecS}_{\vecn_2} 
- g\mu_{\rm B}B\sum_{\vecf\in C}\hat{S}_{\vecf}^z 
\; ,
\label{eq:Heisenbergmodelcluster}
\end{equation}
where the sums run over all lattice sites $\vecf$ in cluster~$C$,
containing $n_C=1\ldots{}4$ spins.
We include the interaction with the external field~$B$
where $g=2$ is the gyromagnetic ratio 
and $\mu_{\rm B}$ is the Bohr magneton.
For our comparisons with experiment, 
we use the experimental values for $J_n$ from table~\ref{tab:Jvalues}
and theoretical values from the GGA+Gutzwiller approach.
However, the differences
between scaled GGA, GGA+$U$, and GGA+Gutzwiller are fairly small.

For each cluster~$C$, we determine its contribution
to the magnetization per lattice site,
\begin{equation}
M^C(B)= \frac{1}{n_C}{\rm Tr}\Bigl(\hat{\rho}_C \sum_{\vecf\in C}
\hat{S}_{\vecf}^z \Bigr)
\;, \;
\hat{\rho}_C = \frac{e^{-\beta \hat{H}^C_{\rm Heis}(B)}}{
{\rm Tr}\bigl( e^{-\beta \hat{H}^C_{\rm Heis}(B)}\bigr)}
\end{equation}
with $\beta=1/(k_{\rm B}T_{\rm spin})$. The trace is readily calculated
using the exact spectrum 
that we obtain from a complete diagonalization
of the cluster Hamiltonian $\hat{H}^C_{\rm Heis}(B)$.
The magnetization per lattice site is then given by
the sum over all clusters weighted by their multiplicity~$A_C$ and probability~$p_C$,
\begin{equation}
M(B)= \sum_C A_C{} p_C M^C(B) \; .
\end{equation}
We show the resulting magnetization in Fig.~\ref{fig:bindilattimag}.
\pagebreak[3]

The curve for zero temperature shows the expected magnetization steps
that occur when more and more Mn pairs (or clusters) 
align with the external field.~\cite{Foner1989} 
When we use the experimentally determined values for the
exchange couplings from table~\ref{tab:Jvalues}
and a spin temperature~$T_{\rm spin}=100\, {\rm mK}$
that is somewhat higher than the environment 
temperature~$T=20\, {\rm mK}$,~\cite{Bindilatti2000}
we find that the agreement between theory and experiment 
for $M(B)$ is very good.
The agreement becomes slightly worse 
when we use the coupling parameters
calculated by GGA+Gutzwiller. Note that the experimentally
accessible magnetic fields probe
mostly $J_2$, $J_3$ and $J_4$ because we have 
$k_{\rm B}T,g\mu_{\rm B}B\ll J_1$ 
and $k_{\rm B}T,g\mu_{\rm B}B \gg J_{n\geq 5}$.

\subsection{Density and spin distributions}
\label{sec:localocc}

To gain further insight into the nature of the ground state
of the Mn ion, we present results for the local occupancies.

\begin{figure}[ht]
\begin{center}
\includegraphics[width=1.0\columnwidth]{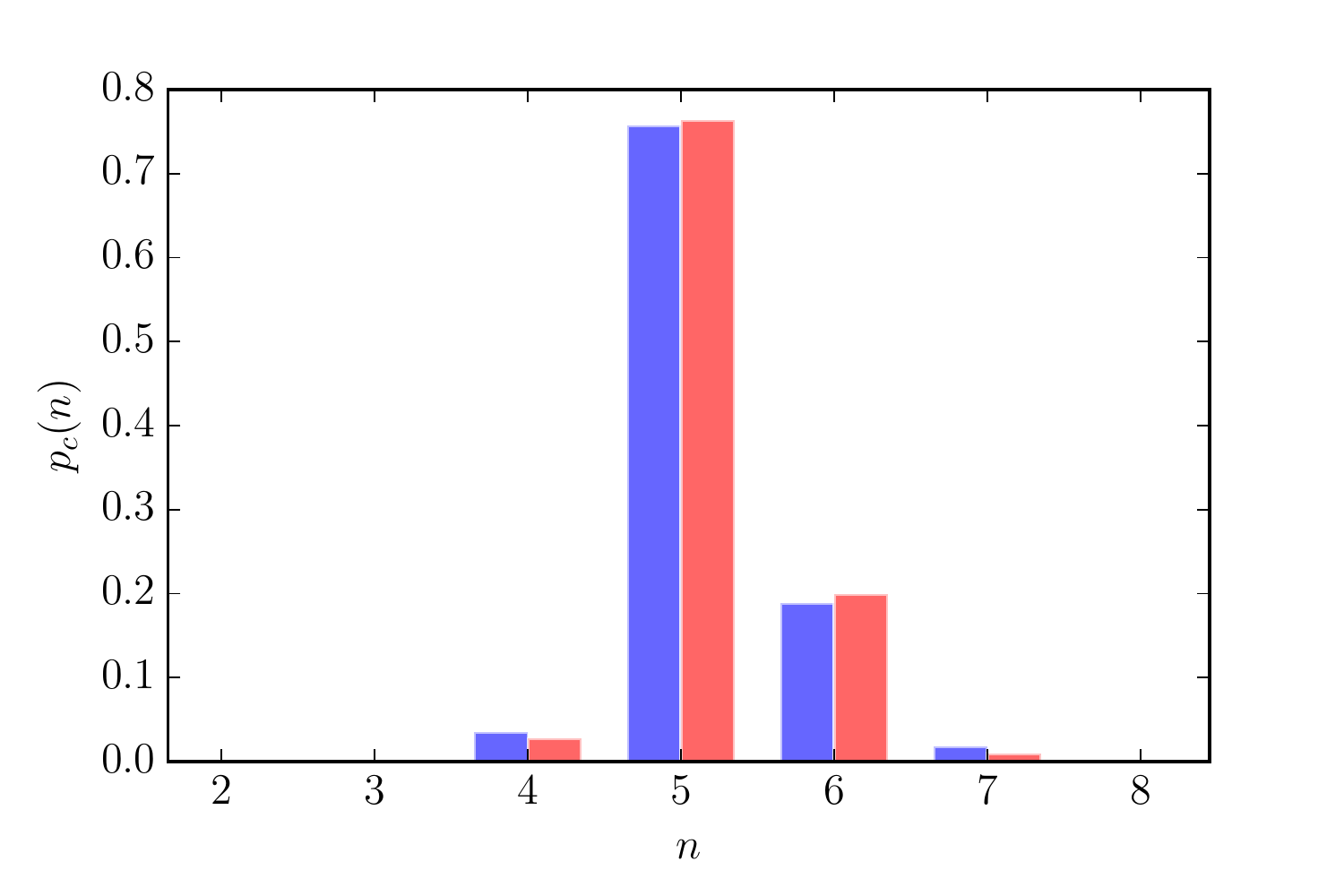}
\caption{Mn charge distribution~$p_c(n)$ as a function of the number~$n$ 
of $3d$-electrons for $A=4.4\,{\rm eV}$, $B=0.1\,{\rm eV}$, $C=0.4\,{\rm eV}$ 
for $n_d=5.19$ in GGA+Gutzwiller
(red columns), in comparison with the Hartree--Fock result (blue columns).
\label{fig:nd-distribution}}
\end{center}
\end{figure}

We start our discussion with the probability distribution $p_c(n)$ to find 
$n$~$3d$~electrons on the Mn ion on site~$\vecf$ ($0\leq n\leq 10$).
As seen from Fig.~\ref{fig:nd-distribution} the distribution peaks
at $n=5$ which reflects 
the fact that the average particle number is $n_d=5.19$,
see Sect.~\ref{sec:couplingstrengths}.
Correspondingly, there also is a sizable probability
to find $3d^6$ configurations on the Mn ion whereas 
the probability for all other occupation numbers is negligible.
Note that this distribution 
is not the result of electronic correlations 
because the corresponding Hartree--Fock
state displays almost the same distribution function.

\begin{figure}[ht]
\begin{center}
\includegraphics[width=1.0\columnwidth]{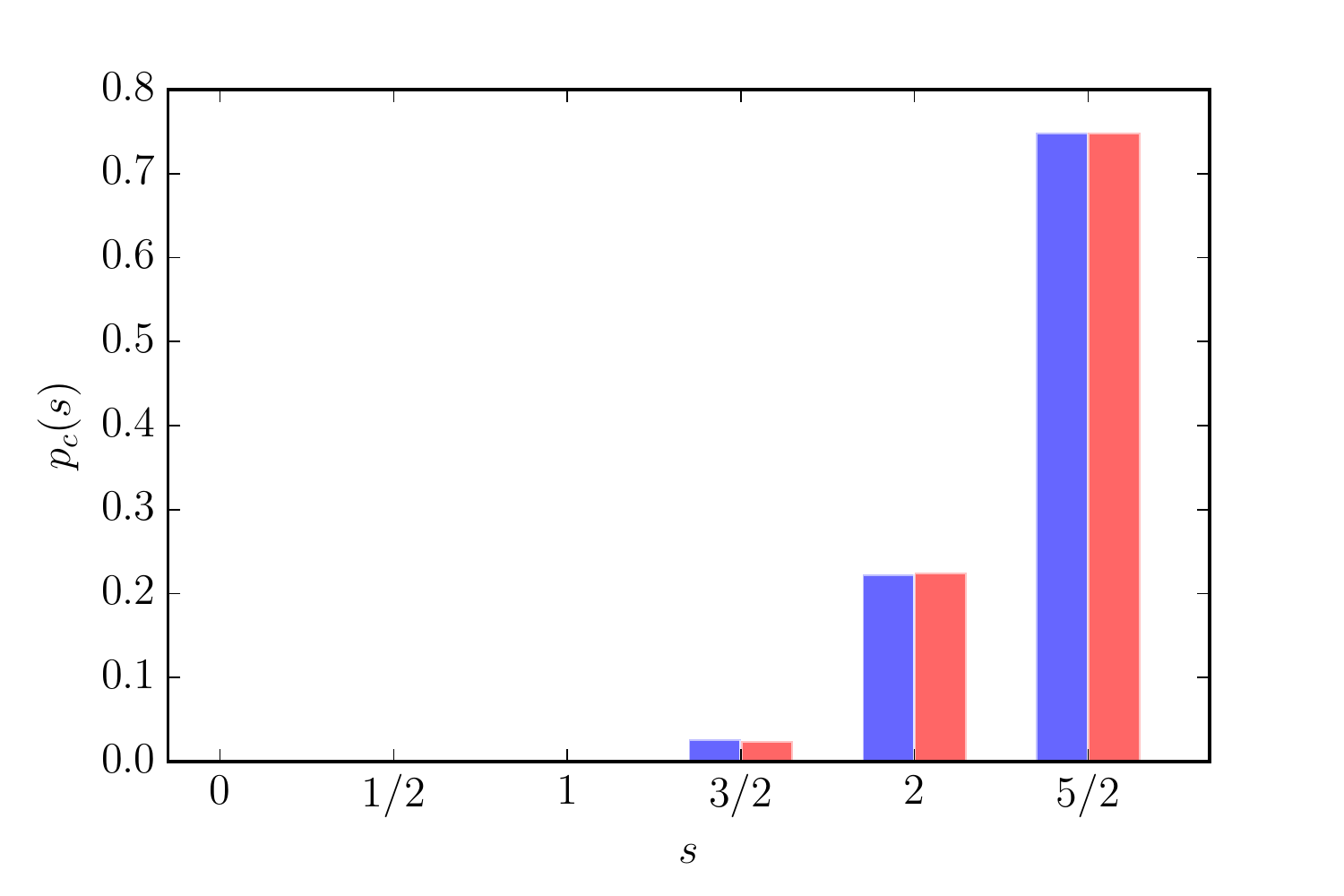}
\caption{Mn spin distribution~$p_s(s)$ as a function of the spin~$s$ 
of $3d$-electrons for  $A=4.4\,{\rm eV}$, $B=0.1\,{\rm eV}$, $C=0.4\,{\rm eV}$ 
for $n_d=5.19$ in GGA+Gutzwiller
(red columns), in comparison with the Hartree--Fock result (blue columns).
\label{fig:ns-distribution}}
\end{center}
\end{figure}

The probability distribution function $p_s(s)$ for finding local spins with size
$0\leq s\leq 5/2$ is very similar to the distribution in the single-particle 
product state~$|\Phi_0\rangle$, i.e., 
the correlation enhancement of the local spin moment is also small
for the spin distribution function, see Fig.~\ref{fig:ns-distribution}.
The average local spin is $\langle \hat{S}_{\vecf}^z\rangle_{\rm loc}=2.33$ because
the admixture of spin $s=2$ to the dominant configurations with $s=5/2$
is not negligibly small. 

The Mn ions do not show integer filling $n_d=5$, nor does the spin moment
correspond to the atomic spin $s=5/2$. 
This observation puts into question the concept of a Heisenberg-model
description that we employed in Sect.~\ref{sec:heismodel} to derive
the exchange couplings.
Even if we accept a non-integer filling of the Mn ions' $3d$-shell, 
we are actually far from a local-moment regime
that is implicit in the Heisenberg-model description~(\ref{eq:Heisenbergmodel})
in Sect.~\ref{sec:heismodel}. This issue can be resolved as seen in the 
next section.

\section{Magnetic response of ion pairs at non-integer filling}
\label{sec:toymodels}

In order to reconcile the finding of a non-integer Mn $3d$
filling and the notion of a spin $s=5/2$ effective Heisenberg model,
we study the magnetic response in a simplified toy model of two Mn
atoms close to their Hund's rule ground states that are coupled 
to three uncorrelated sites.
The uncorrelated sites serve two purposes, namely,
(i), they act as a reservoir to adjust the average particle
number on the Mn sites away from integer filling
and, (ii), they serve as an intermediate charge-transfer 
(exchange) site to mimic the super-exchange mechanism.

\subsection{Model Hamiltonian}

\begin{figure}[tb]
\begin{center}
\includegraphics[width=0.9\columnwidth]{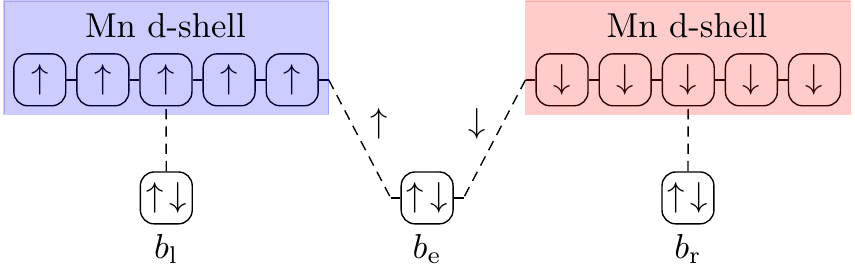}
\caption{Toy model for the study of two Mn atoms with
two reservoir sites and an indirect Mn-Mn coupling via a charge-transfer site.
\label{fig:twotoymodels}}
\end{center}
\end{figure}

The Hamiltonian for our few-site toy-model, illustrated
in Fig.~\ref{fig:twotoymodels}, is readily formulated. 
We use the local Hamiltonian $\hat{H}_{\vecg}^{\rm loc}$ 
defined in eq.~(\ref{eq:Hamiltonian}) for the Mn atoms at $\vecf_{\rm l}$ 
and $\vecf_{\rm r}$,
and 
\begin{equation}
\hat{H}_{\rm j}^{\rm loc} 
= \epsilon_{\rm j} \sum_{\sigma} \hat{n}_{{\rm j},\sigma}  \; ,
\end{equation}
for the local Hamiltonians of the three uncorrelated orbitals.
Here, $\epsilon_{\rm j}$ (${\rm j}={\rm l},{\rm e}, {\rm r}$)
are the local chemical potentials that permit 
the adjustment of the average electron number 
in the left~(l) and right~(r) bath orbitals and the exchange~(e) orbital,
and $\hat{n}_{{\rm j},\sigma} =\hat{c}_{{\rm j},\sigma}^{\dagger}
\hat{c}_{{\rm j},\sigma}^{\vphantom{\dagger}}$ counts the number of electrons
in the uncorrelated orbitals.
The sites are coupled via the kinetic terms
\begin{eqnarray}
\hat{T}_{{\rm l/e},\vecf_{\rm l}}&=&
\sum_{c,\sigma} T_{\vecf_{\rm l},c,\sigma}^{{\rm l/e},\sigma}
\hat{c}_{\vecf_{\rm l},c\sigma}^{\dagger}\hat{c}_{{\rm l/e},\sigma}^{\vphantom{\dagger}} 
+\hbox{h.c.}\; ,\nonumber 
\\
\hat{T}_{{\rm e/r},\vecf_{\rm r}} &=&
\sum_{c,\sigma} T_{\vecf_{\rm r},c,\sigma}^{{\rm e/r},\sigma}
\hat{c}_{\vecf_{\rm r},c\sigma}^{\dagger}
\hat{c}_{{\rm e/r},\sigma}^{\vphantom{\dagger}} +\hbox{h.c.}\;.
\label{eq:Tfortoy}
\end{eqnarray}
The full model Hamiltonian reads
\begin{eqnarray}
\hat{H}&=& \hat{H}_{\vecf_{\rm l}}^{\rm loc} + \hat{H}_{\vecf_{\rm r}}^{\rm loc} 
+ \hat{H}_{\rm l}^{\rm loc} + \hat{H}_{\rm e}^{\rm loc} 
+ \hat{H}_{\rm r}^{\rm loc} 
\nonumber \\
&&
+ \hat{T}_{{\rm l},\vecf_{\rm l}}+ \hat{T}_{\rm e,\vecf_{\rm l}}+ \hat{T}_{\rm e,\vecf_{\rm r}}
+  \hat{T}_{{\rm r},\vecf_{\rm r}} \; .
\end{eqnarray}
The maximal dimension of the corresponding
Fock space is ${\rm dim}H=1024^2\cdot 4^3$.
It is  too large to be handled exactly.

{}From our analysis 
in Sect.~\ref{sec:results} we know that, 
for large $U$, $J$, those Mn configurations are dominantly occupied
that, in the sectors with $n_d=4,5,6$ electrons,
have maximal spin $s_{\rm max}=2,5/2,2$ 
and maximal orbital momentum $l_{\rm max}=2,0,2$, 
which is a good quantum number in spherical approximation.
Therefore, we restrict the Hilbert space of our two Mn atoms 
to these atomic subspaces. To this end, we
introduce the projection operators ${\cal P}_{\vecg,n_d}^{\rm H}$ 
onto the lowest-lying $(2s_{\rm max}+1)(2l_{\rm max}+1)$
Hund's-rule states for fixed electron number $n_d$,
\begin{eqnarray}
\hat{H}_{\vecg}^{\rm loc}|\Gamma_n\rangle_{\vecg}
&=&E_n^{\rm loc}|\Gamma_n\rangle_{\vecg}\;,  \nonumber \\
\hat{n}_{\vecg}|\Gamma_n\rangle_{\vecg}
&=&n_d|\Gamma_n\rangle_{\vecg}\; , \nonumber \\
\hat{\vecS}_{\vecg}^2|\Gamma_n\rangle_{\vecg}
&=&s_{\rm max}(s_{\rm max}+1)|\Gamma_n\rangle_{\vecg}\; , \nonumber \\
\hat{\vecL}_{\vecg}^2|\Gamma_n\rangle_{\vecg}
&=&l_{\rm max}(l_{\rm max}+1)|\Gamma_n\rangle_{\vecg}\; , \nonumber \\
{\cal P}_{\vecg,n_d}^{\rm H}&=& \sum_{\Gamma_n} 
|\Gamma_n\rangle_{\vecg}
\,{}_{\vecg}\langle\Gamma_n| \, .
\end{eqnarray}
Then, we define the total projection operator
\begin{equation}
{\cal P}_{4,5,6}^{\rm H}=\left(\sum_{n_d=4}^{6}{\cal P}_{\vecf_{\rm l}, n_d}^{\rm H}\right)
\left(\sum_{n_d=4}^{6}{\cal P}_{\vecf_{\rm r}, n_d}^{\rm H}\right)\; ,
\end{equation}
and we limit ourselves to the investigation of our model Hamiltonians in the 
projected form
\begin{equation}
{\cal H} = {\cal P}_{4,5,6}^{\rm H} \hat{H} {\cal P}_{4,5,6}^{\rm H} \; .
\end{equation}
The dimension of the partial Fock space on the Mn atoms
is $(2s_{\rm max}+1)(2l_{\rm max}+1)$
so that the maximal Fock-space dimension is
${\rm dim}{\cal H} = (25+6+25)^2\cdot 4^3=200704$.
This partial Fock space is accessible using the Lanczos technique.

\subsection{Magnetization plateaus}

The magnetic field couples to the spin-component 
of the Mn atoms in $z$-direction,
\begin{equation}
\hat{H}_B=-g\mu_{\rm B}B(\hat{S}_{\vecf_{\rm l}}^z+\hat{S}_{\vecf_{\rm r}}^z) \; .
\end{equation}
The magnetization is obtained from
\begin{equation}
M(B)=\langle \Psi_0 | \hat{S}_{\vecf_{\rm l}}^z+\hat{S}_{\vecf_{\rm r}}^z | \Psi_0\rangle
\;, 
\end{equation}
where 
$|\Psi_0\rangle$ is the ground-state of our model Hamiltonian
in the presence of a magnetic field,
\begin{equation}
{\cal H}(B)={\cal H}+\hat{H}_B \; .
\end{equation}
We employ the Lanczos algorithm to find $|\Psi_0(B)\rangle$.

We fix the total number of electrons in the system to $n_{\rm tot}=16$,
and choose the local chemical potentials $\epsilon_{\rm l}= \epsilon_{\rm r}$ 
to adjust the average electron number on the Mn sites
so that we have an average number of $n_d=5.30$ electrons.
Note that this number marginally changes as a function of the magnetic field.
We set all electron transfer matrix equal in eq.~(\ref{eq:Tfortoy}),
$T_{..}^{..}=1\, {\rm eV}$.

\begin{figure}[t]
\begin{center}
\includegraphics[width=0.9\columnwidth]{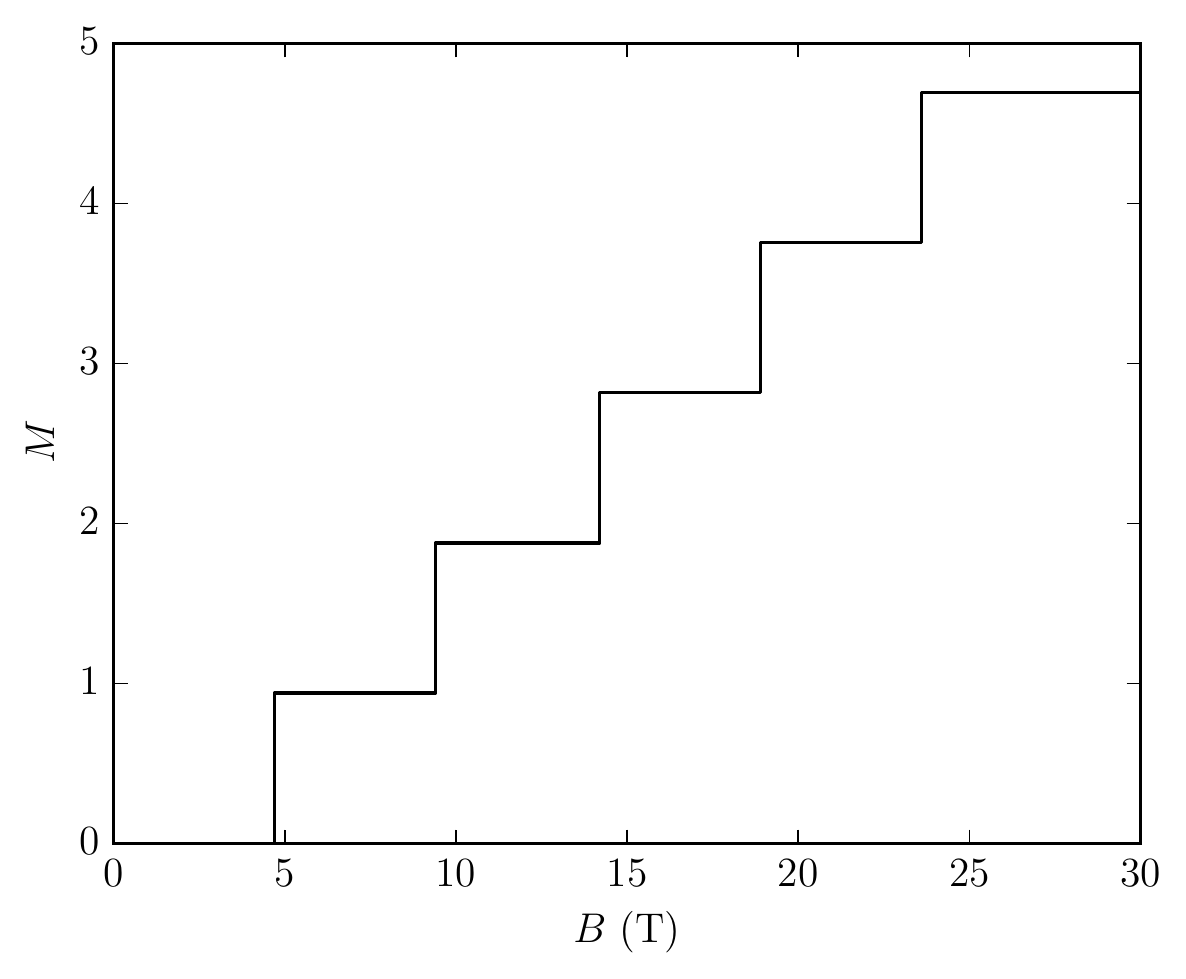}
\caption{Magnetization $M(B)$ as a function of the external field for
the toy model with an almost filled exchange site,
$n_{\rm e}=1.98$.\label{fig:equidistantsteps}}
\end{center}
\end{figure}

In the following case~(i), we set $\epsilon_{\rm e}=8.0\, {\rm eV}$ 
and $\epsilon_{\rm l,r}=23.1\, {\rm eV}$
so that we have $n_{\rm e}=1.98$ electrons in the exchange orbital
and $n_{\rm l,r}=1.71$ electrons in each bath orbital in the ground state.
The resulting magnetization steps are equidistant, as shown 
in Fig.~\ref{fig:equidistantsteps},
despite the fact that the Mn filling is far from integer.

\begin{figure}[t]
\begin{center}
\includegraphics[width=0.9\columnwidth]{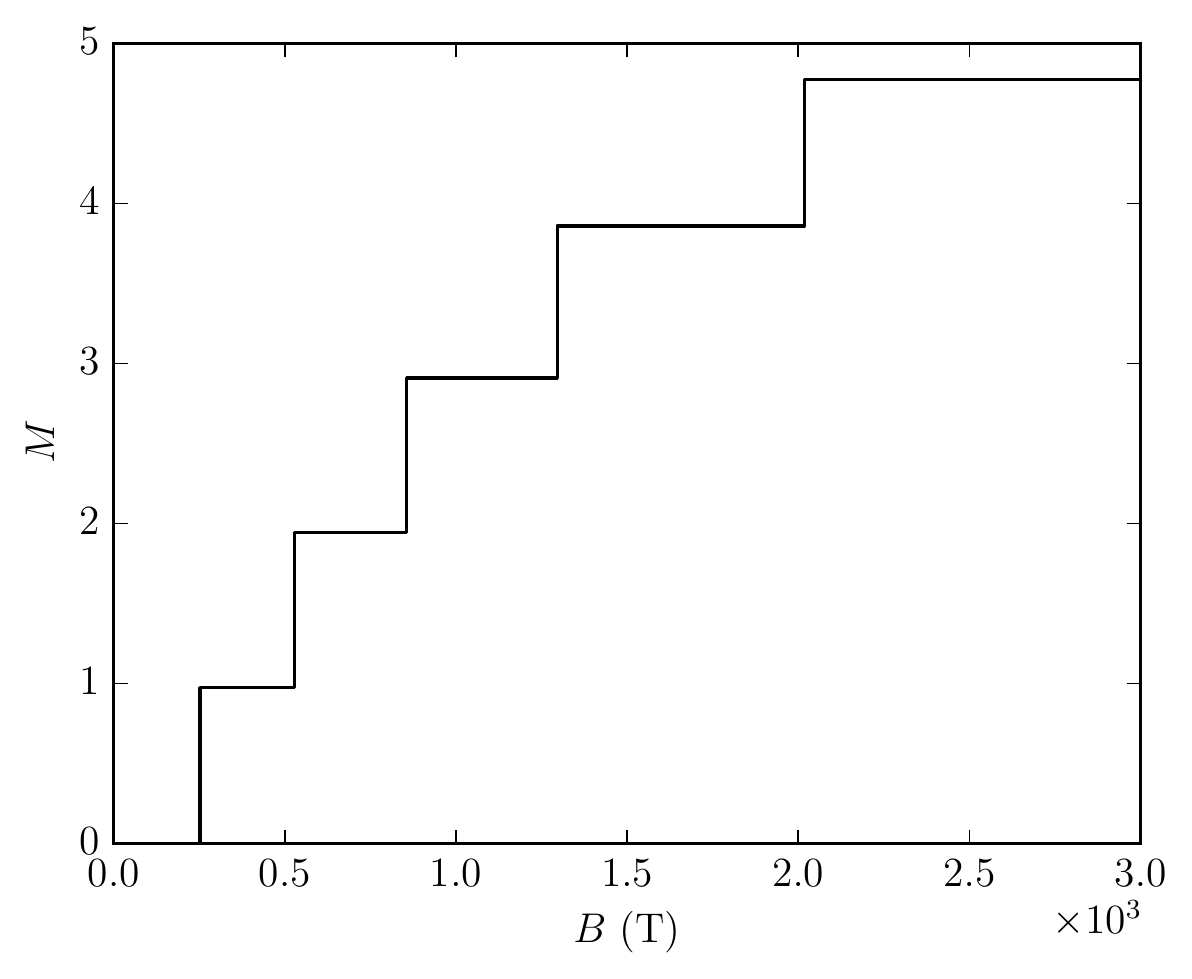}
\caption{Magnetization $M(B)$ as a function of the external field for
the toy model with a partly filled 
exchange site, $n_{\rm e}=1.51$.\label{fig:inequivalentsteps}}
\end{center}
\end{figure}

The width of the magnetization steps become non-uniform in case~(ii)
when the exchange site is not almost filled.
To illustrate this case, we choose $\epsilon_{\rm e}=22.6\, {\rm eV}$ 
and $\epsilon_{\rm l,r}=19.5\, {\rm eV}$
so that we have $n_{\rm l,r}=1.93$ electrons in each bath orbital and
$n_{\rm e}=1.51$ electrons in the exchange orbital.
Now, the lengths 
of the corresponding magnetization plateaus are inequivalent,
as shown in Fig.~\ref{fig:inequivalentsteps}.

The toy model shows that equidistant plateaus are possible even
though the occupation of the Mn sites is not integer.
Our numerical observations can be readily understood
using perturbative arguments.
For negligible couplings to the exchange orbital,
the ground state of each Mn ion and its attached bath site 
has spin $s=5/2$. Note that this spin is not solely located
on the Mn site but also partly on the corresponding bath site.
In case~(i), 
the exchange orbital introduces only a small coupling
between the left and the right spin-5/2 systems, and
perturbation theory leads to a dominant term of the usual 
antiferromagnetic Heisenberg form~(\ref{eq:Heisenbergmodel}).
Consequently, the magnetization steps are 
equidistant.~\cite{PhysRevB.33.1789,Foner1989}
In case~(ii), 
charge fluctuation contributions 
invalidate the simple spin-only picture.
This results in non-equidistant magnetization steps as
seen in Fig.~\ref{fig:inequivalentsteps}.

When we apply the Gutzwiller approximation scheme used 
in Sect.~\ref{sec:results} to case~(i) of our toy model, we find
an exchange coupling~$J_1$ that is very close to
the exact value derived from the width of the magnetization plateaus.
This corroborates our finding in Sect.~\ref{sec:results} and further justifies
the applicability of our toy model.

Due to the large gap for charge excitations,
the situation of Mn ions in CdTe resembles scenario~(i)
in our toy model and explains the experimental observation of equidistant
magnetization plateaus. The filling
of the Mn $3d$-shell is not integer but the total spin of the
Mn ion and its surrounding atoms still is essentially $s=5/2$.

\section{Conclusions}
\label{sec:conclusions}

In this work we used three band structure methods, DFT(GGA),
GGA+$U$, and GGA+Gutzwiller, to derive the exchange couplings between Mn 
ions
diluted in II-VI semiconductor host materials such as CdTe.
First, we calculate the energy of the configurations
with parallel and antiparallel alignments of the Mn spins.
Next, we interpret the energy difference in terms of
a two-spin Heisenberg model and thereby deduce the exchange couplings
as a function of the Mn-Mn separation for up to fourth neighbor distances.

For the GGA calculations we employ the FLEUR code with the functional of 
Perdew,
Burke, and Ernzerhof for large supercells with $L=128$ atoms
where two of the Cd ions are replaced by isovalent Mn ions.
The ab-initio results for the exchange couplings are too large by a 
factor of two to three
which is related to the fact that DFT(GGA) underestimates
gaps in II-VI semiconductors systematically.
The nearest-neighbor couplings $J_1$ for Mn ions in II-VI semiconductors
can be reconciled with experiment
by using the GGA+$U$ and GGA+Gutzwiller methods. These methods
employ adjustable parameters that are used to match the experimental 
value for $J_1$
in Cd(Mn)Te. The exchange couplings $J_{2,3,4}$
for 2nd, 3rd, and 4th neighbor distances are then predictions from theory.

In general, the values for $J_{n\geq 2}$ agree qualitatively with 
experiment, i.e.,
band theory recovers $J_4>J_2,J_3$ and $J_{n\geq 5}\lesssim0.01\, {\rm K}$.
However, the values for the couplings do not agree perfectly, i.e., we 
observe
quantitative deviations up to a factor of two.
About the same level of accuracy can be obtained by a simple rescaling
of the DFT(GGA) data that fits the nearest-neighbor coupling $J_1$,
see tables~\ref{tab:Jvalues}, \ref{tab:three}, \ref{tab:two}, and 
\ref{tab:four}.
The bare energy scale in our itinerant-electron description
are of the order of several eV, i.e.,
of the order of $10^5\, {\rm K}$, whereas the exchange couplings $J_{n\geq 2}$
are one Kelvin and below. Therefore, it does not come as a surprise that
the band structure methods reach their accuracy limits.

The notion of exchange couplings and the applicability of the
super-exchange approach hinges on the mapping of the
low-energy degrees of freedom of the itinerant-electron problem
to those of a spin-5/2 Heisenberg model.
This mapping successfully explains the equidistant magnetization plateaus
as a function of applied magnetic field, as seen in experiment.
However, the analysis of the Gutzwiller ground state
for the two-ion Hubbard model shows that
the filling of the Mn $3d$-shell is {\em not\/} integer
which seemingly invalidates the whole concept of localized spins.
The analysis of an exactly solvable few-site toy-model reassures
that an integer filling is not a prerequisite for equidistant
magnetization plateaus.
Due to the hybridization of the Mn $3d$ orbitals with its insulating environment, 
a slightly delocalized spin-5/2 magnetic moment is formed combing 
Mn $3d^5$ and $3d^6$ with neighboring valence band states.
Our picture of an extended spin-5/2 magnetic moment interacting 
with each other reconciles the usage of an effective spin-5/2 Heisenberg model 
to explain the experimentally observed magnetization steps and simultaneously 
a non-integer valence of the Mn 3d shell.

In the case of Mn-doped II-VI semiconductors, the Gutzwiller method and the
Hartree-Fock approach to the two-ion Hubbard model lead to essentially
the same results for an (anti-)ferromagnetic alignment of the Mn spins.
Our preliminary investigations show that this is not the case
for Cr in CdTe where the dopant electrons are more itinerant
than in the case of Mn doping. We observe the same trend for Mn doping of
GaAs and other III-V semiconductors. This observation also indicates
that the Heisenberg mapping is less appropriate in these cases,
and it is advisable to employ a correlated-electron approach
for the description of the magnetic response in GaAs samples
at low Mn doping.

\begin{acknowledgments}
We thank Prof.\ Valdir Bindilatti for providing us with his original 
magnetization data
for Cd(Mn)Te, Ref.~[\onlinecite{Bindilatti2000}], shown in Fig.~\ref{fig:bindilattimag}.
We also have profited from fruitful discussions with M. Bayer and D. Yakovlev.
We are particularly indebted to Stefan Bl\"ugel for information on
the FLEUR program and for bringing some special aspects of DFT+$U$ to
our attention. Finally, we thank the late Werner Weber
who has initiated this project.

Some of us (T.L., U.L., and F.B.A.) acknowledge the financial support 
by the Deutsche Forschungsgemeinschaft and the Russian Foundation of Basic
Research in the frame of the ICRC TRR 160.

The authors gratefully acknowledge the computing time granted
by the John-von-Neumann Institute for Computing (NIC),
and provided on the supercomputer JURECA at J\"ulich Supercomputing 
Centre (JSC) under project no.\ HDO08.
\end{acknowledgments}


\end{document}